\documentclass[11pt,a4paper]{article}
\usepackage{jheppub_kim}
\usepackage{rotating} 
\usepackage{graphicx,epsfig}
\usepackage{amsmath}
\usepackage {amssymb}
\usepackage{subfigure}

\usepackage{relsize}

\providecommand{\U}[1]{\protect\rule{.1in}{.1in}}

\newcommand{\newc}{\newcommand}

\newc{\be}{\begin{equation}}
\newc{\ee}{\end{equation}}

\newc{\ba}{\begin{eqnarray}}
\newc{\ea}{\end{eqnarray}}
\newc{\bea}{\begin{eqnarray*}}
\newc{\eea}{\end{eqnarray*}}
\newc{\D}{\partial}
\newc{\ie}{{\it i.e.} }
\newc{\eg}{{\it e.g.} }
\newc{\etc}{{\it etc.} }
\newc{\etal}{{\it et al.}}
\newc{\lcdm}{$\Lambda$CDM }

\newc{\ra}{\Rightarrow}

\title{Correspondence between Myrzakulov $F(R,Q)$ gravity and Tsallis 
cosmology  }

 \author[a]{Andreas Lymperis}
 
  \author[b]{Gulgassyl Nugmanova}
   \author[b]{Almira Sergazina}

\affiliation[a]{Department of Physics, University of Patras, 26500 Patras, 
Greece}
\affiliation[b]{Eurasian National University, Astana, 010008, Kazakhstan}

\emailAdd{alymperis@upatras.gr}

\abstract{   
We investigate the correspondence between Myrzakulov $F(R,Q)$ 
gravity  and Tsallis cosmology. The former is a modified gravity that uses 
both 
curvature and nonmetricity, while the latter is a modified cosmology arising 
from the gravity-thermodynamics conjecture,  employing Tsallis entropy instead 
of the   Bekenstein-Hawking one. By appropriately identifying the 
functional dependencies  and the model parameters, we 
demonstrate that both frameworks can give identical background evolution, 
reproducing the standard cosmological sequence of matter and dark energy 
domination. However, their perturbation behavior exhibits  differences, 
since the growth 
of density fluctuations and the effective Newton constant  deviate
between the two scenarios, indicating that perturbative observables, such as 
structure formation and weak-lensing ones, could serve as distinguishing 
factors between them.
}


\begin{document}
\maketitle

\section{Introduction}

 The Standard Model of Cosmology, which consists of $\Lambda$-Cold Dark Matter 
($\Lambda$CDM)  within general relativity, 
has proven to be very effective in explaining the evolution of the 
universe, both in terms of background dynamics and perturbation theory 
\cite{Peebles:2002gy}. Despite its successes, certain theoretical challenges, 
such 
as the cosmological constant problem \cite{Weinberg:1988cp}, the 
non-renormalizability of general 
relativity \cite{Addazi:2021xuf}, and possible observational tensions 
\cite{Abdalla:2022yfr}, have motivated the exploration 
of various extensions and alternative approaches. Broadly speaking, these 
extensions fall into two main categories. The first category retains general 
relativity as the fundamental gravitational framework but introduces additional 
components, such as the dark energy sector  \cite{Copeland:2006wr,Cai:2009zp}. 
The second category involves modifications of gravity itself, where general 
relativity emerges as a limiting case,   capable of driving cosmic acceleration 
and richer cosmological behavior 
\cite{CANTATA:2021asi,Capozziello:2011et,Cai:2015emx}.

 One approach to construct modified gravity theories  
involves modifying the Einstein-Hilbert action by introducing additional terms, 
leading to various extensions such as $f(R)$ gravity 
\cite{DeFelice:2010aj,Nojiri:2010wj,Starobinsky:2007hu} or Gauss-Bonnet and 
$f(G)$ gravity \cite{Nojiri:2005jg,DeFelice:2008wz,Zhao:2012vta,Shamir:2020ckh}.
Alternatively, modifications can arise within the torsional gravitational 
framework, giving rise to theories such as $f(T)$ gravity 
\cite{Bengochea:2008gz,Linder:2010py,Chen:2010va,Tamanini:2012hg,
Bengochea:2010sg,Liu:2012fk,Daouda:2012nj}, as well as its extensions like 
$f(T,T_G)$ gravity \cite{Kofinas:2014owa} and $f(T,B)$ gravity 
\cite{Bahamonde:2015zma}. 
Moreover, there can be other formulations based on nonmetricity and the 
symmetric teleparallel connection, leading to  
$f(Q)$ gravity \cite{BeltranJimenez:2017tkd,Heisenberg:2023lru} and its 
variants 
such as $f(Q,C)$ gravity \cite{De:2023xua}. 

 Within this geometrical framework, Myrzakulov $F(R,Q)$ gravity incorporates 
both curvature and nonmetricity to describe gravitational interactions 
\cite{Myrzakulov:2012qp}. By employing a generalized connection that 
simultaneously accounts for both geometric properties, this theory establishes 
a 
more intricate gravitational structure, leading to a richer  
cosmological behavior. This enhanced framework allows for interesting 
phenomenology, and has led to many applications \cite{Myrzakulov:2012qp, 
Saridakis:2019qwt, Anagnostopoulos:2020lec, Myrzakul:2021kas, 
Iosifidis:2021kqo, 
Myrzakulov:2021vel, Harko:2021tav, Papagiannopoulos:2022ohv, Kazempour:2023kde, 
Zahran:2024nsz, Maurya:2024btu, Maurya:2024nxx, Maurya:2024ign, Maurya:2024mnb, 
Momeni:2024bhm,Lymperis:2025fbs}.

On the other hand, there is another approach to account for the theoretical and 
observational puzzles, namely to apply the gravity-thermodynamics conjecture 
 \cite{Jacobson:1995ab,Padmanabhan:2003gd,Padmanabhan:2009vy}. Specifically, it 
has been demonstrated that the cosmological Friedmann equations can be 
reformulated as the first law of thermodynamics when treating the universe as a 
thermodynamic system bounded by the apparent horizon 
\cite{Frolov:2002va,Cai:2005ra,Akbar:2006kj,Paranjape:2006ca,
Akbar:2006er, Sheykhi:2007zp,
 Cai:2009ph, 
Wang:2009zv,  Gim:2014nba, 
Fan:2014ala,Tsilioukas:2024seh}. Apart from the basic 
formulation, using the standard Bekenstein-Hawking entropy, this approach has 
been applied to extended entropies too, such as Tsallis, Barrow, Kaniadakis etc 
entropies, and  proves to exhibit rich cosmological phenomenology
\cite{Lymperis:2018iuz,Arias:2019zug,Nojiri:2019skr,Geng:2019shx,
Saridakis:2020lrg, Lymperis:2021qty, 
  Sheykhi:2022gzb,DiGennaro:2022ykp,Luciano:2022ely, 
 Coker:2023yxr,Basilakos:2023kvk, 
 Salehi:2023zqg,Jizba:2023fkp,Jizba:2024klq,Ebrahimi:2024zrk,
Tyagi:2024cqp,Okcu:2024llu,Petronikolou:2024zcj,Mendoza-Martinez:2024gbx,
Yarahmadi:2024lzd,Tsilioukas:2025dmy}.  
 
In this paper we examine the Correspondence between Myrzakulov $F(R,Q)$ gravity 
and Tsallis cosmology. In particular, we are interested in studying under which 
connection choices one obtains equivalence of the two frameworks. 
The paper is structured as follows: In Section \ref{section2}, we briefly 
review Myrzakulov $F(R,Q)$ gravity and cosmology, while in Section 
\ref{Tsallis} we present the modified Friedmann equations in Tsallis cosmology. 
Then, in Section \ref{Correspondence} we reconstruct the involved functions and 
parameters in order to make the two theories coincide at the background level, 
and additionally we study the perturbation evolution showing that the two 
theories deviate. Finally, 
we summarize our results  in Section 
\ref{Conclusions}.

\section{Myrzakulov $F(R,Q)$ Cosmology}
\label{section2}

The central idea of this modified gravity is 
the modification of the underlying connection. In particular, we introduce
    a connection    \cite{Maurya:2024ign}
\begin{equation}\label{connectdef}
	\Gamma^{\rho}_{\,\,\,\,\mu\nu}=\hat{\Gamma}_{\,\,\,\, \mu 
\nu}^{\rho}+L^{\rho}_{\,\,\,\,\mu\nu},
\end{equation}
with  $\hat{\Gamma}_{\,\,\,\, \mu \nu}^{\rho}$   the Levi-Civita 
connection and $L^{\rho}_{\,\,\,\,\mu\nu}$    the disformation tensor given by
\begin{eqnarray}
    	L^{\rho}_{\,\,\,\,\mu\nu}&=&\frac{1}{2}g^{\rho\lambda}\bigl(-Q_{\mu \nu 
\lambda}-Q_{\nu \mu \lambda} + Q_{\lambda\mu\nu}\bigr) \label{Ldef},
\end{eqnarray}
and where $Q_{\rho \mu \nu} = \nabla_{\rho} g_{\mu \nu}$ is the nonmetricity 
tensor.

As it is known,  the Ricci curvature tensor $R_{\mu\nu}$ in terms of 
the general connection is written as 
\cite{Harko:2021tav,Capozziello:2022zzh,Shimada:2018lnm} 
\begin{equation}
R_{\mu\nu}=\partial_{\lambda}\Gamma^{\lambda}_{\mu\nu}-\partial_{\mu}\Gamma^{
\lambda}_{\lambda\nu}+\Gamma^{\lambda}_{\lambda\alpha}\Gamma^{\alpha}_{\mu\nu}
-\Gamma^{\lambda}_{\mu\alpha}\Gamma^{\alpha}_{\lambda\nu}, \label{2.4}
\end{equation}
which can be expressed in terms of the Levi-Civita connection as
\begin{equation}
R_{\mu\nu}=\hat{R}_{\mu\nu}+\partial_{\lambda}L^{\lambda}_{\mu\nu}-\partial_{
\mu}L^{\lambda}_{\lambda\nu}+\hat{\Gamma}^{\lambda}_{\lambda\alpha}L^{\alpha}_
{\mu\nu}+\hat{\Gamma}^{\alpha}_{\mu\nu}L^{\lambda}_{\lambda\alpha}-\hat{
\Gamma}^{\lambda}_{\mu\alpha}L^{\alpha}_{\lambda\nu}-\hat{\Gamma}^{\alpha}_{
\lambda\nu}L^{\lambda}_{\mu\alpha}+L^{\lambda}_{\lambda\alpha}L^{\alpha}_{\mu\nu
}-L^{\lambda}_{\mu\alpha}L^{\alpha}_{\lambda\nu}, \label{2.5}
\end{equation}
where by $\hat{R}_{\mu\nu}$ we mark the Ricci curvature tensor with 
respect to Levi-Civita connection. Hence, 
 the Ricci scalar $
 R=g^{\mu\nu}R_{\mu\nu}$
  with respect to the general    
connection  can be re-written  as
\begin{equation}
	R=\hat{R}+u,  
\end{equation}
where $u $ is a   function   depending on 
the   metric, its first and second derivatives, and the connection 
and its first derivative.

In the same lines, the nonmetricity tensor   with 
respect to general     connection  can be expressed as
\begin{equation}	
Q_{\rho\mu\nu}=\partial_{\rho}g_{\mu\nu}-\Gamma^{\lambda}_{\mu\rho}g_{\lambda\nu
}-\Gamma^{\lambda}_{\nu\rho}g_{\lambda\mu},  
\end{equation}
which can be re-written as
\begin{equation}	
Q_{\rho\mu\nu}=\breve{Q}_{\rho\mu\nu}+(-L^{\lambda}_{\mu\rho}g_{\lambda\nu}-L^{
\lambda}_{\nu\rho}g_{\lambda\mu}),  
\end{equation}
where $\breve{Q}_{\rho\mu\nu}$ is the nonmetricity scalar calculated with the 
symmetric teleparallel connection.
Thus, we can express the nonmetricity scalar $ 
Q= 
-g^{\mu\nu}(L^{\alpha}_{\beta\mu}L^{\beta}_{\nu\alpha}-L^{\alpha}_{\beta\alpha}
L^{\beta}_{\mu\nu})$ 
 as
\begin{equation}
	Q=\breve{Q}+w,  
\end{equation}
where $w $ is a   function depending   on the metric and its first 
derivative and on the connection.

One can now proceed to introducing the action of the theory as   
\cite{Maurya:2024ign}
 \begin{equation}\label{actiongen}
 S=\frac{1}{16\pi G}\int \left[F(R, Q)+  L_m\right]\sqrt{-g}~d^4x,
 \end{equation}
where $g$ is the metric determinant,   
$  G$ is  the 
Newton constant, and $L_m$ is the matter Lagrangian. 
 Thus, performing variation in terms of the metric yields \cite{Maurya:2024ign}
\begin{equation}\label{3.5}
- \frac{1}{2} g_{\mu \nu} F + F_R R_{(\mu \nu)} + F_Q L_{(\mu \nu)} + 
\hat{\nabla}_\lambda \left(F_Q {J^\lambda}_{(\mu \nu)} \right) + g_{\mu \nu} 
\hat{\nabla}_\lambda \left(F_Q \zeta^\lambda \right) = 8\pi G T_{\mu \nu} \,,
\end{equation}
 with $F_{R}=\frac{\partial F}{\partial R}$, $F_{Q}=\frac{\partial F}{\partial 
Q}$ and where as usual the  matter energy-momentum tensor is defined as $T_{\mu 
\nu }\equiv -\frac{2}{\sqrt{-g}}\frac{\delta \left( 
\sqrt{-g}\mathcal{L}_{m}\right) }{\delta g^{\mu \nu }}$.
In the above expressions we have introduced the quantities
\begin{eqnarray} 
&&L_{\mu \nu}  =\frac{1}{4} \left[ \left(Q_{\mu \alpha \beta} - 2 Q_{\alpha 
\beta \mu} \right) {Q_\nu}^{\alpha \beta} + \left( Q_\mu + 2 \tilde{Q}_\mu 
\right) Q_\nu + \left( 2 Q_{\mu \nu \alpha} - Q_{\alpha \mu \nu} \right) 
Q^\alpha \right]   \\
&&{J^\lambda}_{\mu \nu}  = \sqrt{-g} \left(\frac{1}{4} {Q^\lambda}_{\mu \nu} - 
\frac{1}{2} {Q_{\mu \nu}}^\lambda \right) \,, 
\end{eqnarray}
where  $Q_\lambda=Q_{\lambda \ \ \alpha}^{\ \ \alpha}$ and 
$\tilde{Q}^\lambda=Q_{\alpha}^{\ \lambda \alpha}$, while 
$
\hat{\nabla}_\lambda  = \left( - \frac{1}{2} Q^\lambda +  
\tilde{Q}^\lambda \right)-
  \nabla_\lambda \,, 
$
with   $\nabla_\lambda$ the covariant derivative associated with the general  
connection and $\hat{\nabla}_\lambda $ the covariant derivative corresponding 
to the Levi-Civita connection. 
Additionally, the connection field equations are \cite{Maurya:2024ign}
\begin{equation} 
	{P_\lambda}^{\mu \nu} (F_R) + F_Q \left[ 2 {Q^{[\nu \mu]}}_\lambda - 
{Q_\lambda}^{\mu \nu} + \left( \tilde{Q}^\nu - Q^\nu \right) \delta^\mu_\lambda 
+ Q_\lambda g^{\mu \nu} + \frac{1}{2} Q^\mu \delta^\nu_\lambda \right] = 0 \,,
\end{equation}
where ${P_\lambda}^{\mu \nu} (F_R)$ is  the   Palatini-like tensor given by 
\begin{equation}\label{3.9}
	{P_\lambda}^{\mu \nu} (F_R) := - \frac{\nabla_\lambda \left(\sqrt{-g} F_R 
g^{\mu \nu} \right)}{\sqrt{-g}} + \frac{\nabla_\alpha \left( \sqrt{-g} F_R 
g^{\mu \alpha}\delta^\nu_\lambda \right)}{\sqrt{-g}} \,.
\end{equation}

  We proceed to the cosmological application. We choose a   flat 
Friedmann-Robertson-Walker (FRW) metric, namely
\begin{eqnarray}
 ds^2=- dt^2+a^2(t)\,  \delta_{ij} dx^i dx^j,
 \end{eqnarray}
 where $a(t)$ is the scale factor. In this case, the Ricci scalar calculated in 
terms of the Levi-Civita connection is 
  \begin{eqnarray}
\hat {R}=6\left(\frac{\ddot{a}}{a}+\frac{\dot{a}^{2}}{a^{2}}\right),
\end{eqnarray}
 while the nonmetricity scalar  calculated in 
terms of the symmetric teleparallel connection is 
 \begin{eqnarray}
\breve {Q}&=&6\frac{\dot{a}^{2}}{a^{2}}  .
\end{eqnarray}

In order to extract the field equations we follow the mini-superspace  
approach \cite{Saridakis:2019qwt}  and we set $u=u(a,\dot{a},\ddot{a})$ and 
$w=w(a,\dot{a})$. Moreover, for the matter Lagrangian we set  $L_m=-\rho_m(a)$, 
with $\rho_m$   the 
matter energy density considered to correspond to a perfect fluid
\cite{Paliathanasis:2015aos}.
Thus, the Lagrangian of the action (\ref{actiongen}) becomes
 \begin{eqnarray} 
&&
\!\!\!\!\!\!\!\!\!\!\!\!\!\!\!\!\!\!\!\!\!\!\!\!\!\!\!\!\!\!\!\!\!\!\!\!\!\!\!\!
{\cal L}(a,R,Q,\dot{a},\dot{R},\dot{Q})= 
a^{3}(F-RF_{R}-QF_{Q})+6a\dot{a}^{2}(F_{R}+F_{Q})\nonumber\\
&& \ \ \ \ \ \  \ +6a^{2}\dot{a}\dot{F
} _ { R } +a^ { 3 }
(uF_{R}+wF_{Q})+16\pi G a^{3}L_{m}.
\label{miniLagr}
\end{eqnarray} 
Hence, varying the above Lagrangian provides the field equations 
\begin{eqnarray} 
&&-\frac{1}{2}(F-RF_{R}-QF_{Q})+3H^{2}(F_{R}+F_{Q})-\frac{1}{2}[(u-\dot{a}u_{
\dot {
a}})F_{R}+(w-\dot{a}w_{\dot{a}})F_{Q}]
\nonumber\\
&& 
+3H(\dot{R}F_{RR}+\dot{Q}F_{RQ})=8\pi G 
\rho_m
\end{eqnarray}
and 
\begin{multline} 
\!\!\!\!\!\!\!\!\! 
-\frac{1}{2}(F-RF_{R}-QF_{Q})+(2\dot{H}+3H^{2})(F_{R}+F_{Q})-\frac{1}{2}
(u+\frac 
{1}{3}au_{a}-\dot{a}u_{\dot{a}}-\frac{1}{3}a\dot{u}_{\dot{a}})F_{R}
\\\! \!\!  -\frac{1}{2} 
(w+\frac{1}{3}aw_{a}-\dot{a}w_{\dot{a}}-\frac{1}{3}a\dot{w}_{\dot{a}})F_{Q} 
+2H(\dot{F}_{R}+\dot{F}_{Q})+\frac{1}{6}a(u_{\dot{a}}\dot{F}_{R}+w_{\dot{a}} 
\dot{F}_{Q})+\ddot{F}_{R}=-8\pi G  p_m,
\end{multline}
with   $H=\frac{\dot{a}}{a}$    the Hubble parameter and  $p_m$   the matter 
pressure.
 
 In this paper we will investigate the simplest model, namely 
 \begin{equation} 
 F(R,Q)= R+\lambda Q+\tilde{\Lambda}.
   \end{equation}
   Inserting this ansatz into the FRW field equations we finally extract the 
modified Friedmann equations as  
\begin{eqnarray}
3H^{2}&=& 8\pi G \left( 
\rho_m+\rho_{DE} \right)
\label{FR1a}
\\
2\dot{H}+3H^2
&=&  -8\pi G \left(p_m+ p_{DE}\right),
\label{FR2a}
\end{eqnarray}
 where we 
have defined an effective dark energy sector, with    energy density and 
pressure   respectively given by 	
\begin{eqnarray}
&&
\!\!\!\!\!\!\!\!\!\! 
\rho_{DE}=\frac{1}{8\pi G} 
\left[ \frac{1}{2}(u-\dot{a}u_{\dot{a}})+ \frac{\lambda}{2} 
(w-\dot{a}w_{\dot{a}
}) -3\lambda H^2+\frac{\tilde{\Lambda}}{2} \right]
\label{rhoDEa1}\\
&&
\!\!\!\!\!\!\!\!\!\! 
p_{DE}=
-\frac{1}{8\pi G}
\left[-\frac{1}{2}(u+\frac{1}{3}au_{a}-\dot{a}u_{\dot{a}} 
-\frac{1}{3}a\dot{u}_{\dot{a}})-\frac{ \lambda}{2} 
(w+\frac{1}{3}aw_{a}-\dot{a}w_ {\dot{a}} 
-\frac{1}{3}a\dot{w}_{\dot{a}})\right.\nonumber\\
&&\left. \ \ \ \  \ \ \ \  \ \ \ \  \ \  \,
-\frac{\tilde{\Lambda}}{2}+\lambda(2\dot{H}+3H^{
2})\right],
\label{pDEa1}
\end{eqnarray}
where   the subscripts  
$a,\dot{a},\ddot{a}$ mark the  partial derivatives with respect to them. 
As we can see, assuming that  the matter sector is conserved, we obtain the 
effective dark energy conservation as
 \begin{eqnarray}
 \dot{\rho}_{DE}+3H(\rho_{DE}+p_{DE})=0.
\end{eqnarray}
 Finally, we can introduce the equation-of-state parameter of the effective 
dark energy as $w_{DE}\equiv p_{DE}/\rho_{DE}$.

\section{Tsallis cosmology}
\label{Tsallis}

In this section we   present    Tsallis cosmology, namely the  modified 
Friedmann  equations that arise in the gravity-thermodynamics framework if ones 
uses the extended Tsallis entropy instead of the standard Bekenstein-Hawking 
one.

Tsallis entropy is a generalization of the standard Boltzmann-Gibbs entropy, 
introduced by   Tsallis in 1988 as a framework for extending 
statistical mechanics to non-extensive systems  
\cite{Tsallis:1987eu,Lyra:1997ggy,Tsallis:1998ws,Wilk:1999dr}. It is defined as
\begin{equation}
S_T = k_B \frac{1 - \sum_i p_i^q}{q-1},
\end{equation}
where $ q $ is the entropic index that characterizes the degree of 
non-additivity, $ p_i $ represents the probability of the system's 
microstates, and $ k_B $ is the Boltzmann constant. In the limit $ q \to 1 
$, Tsallis entropy recovers the classical Boltzmann-Gibbs entropy. Tsallis 
entropy   is   relevant in systems exhibiting long-range 
interactions, correlations, or fractal-like phase space structures, where the 
standard thermodynamic formalism fails. 
In the case of a black hole, the 
 nonextensive Tsallis entropy can be written in compact form as 
\cite{Tsallis:2012js}
\begin{equation}
\label{Tsallisentropy}
S_T=\frac{\tilde{\alpha}}{4G} A^{\delta}, 
\end{equation}
with $A\propto L^2$   the area of the system, having characteristic length 
$L$, 
and    $\tilde{\alpha}$ and $\delta$ the parameters. As one can see, 
for $\delta=1$ and $\tilde{\alpha}=(4G)^{-1} $ (in units where $\hbar=k_B = c = 
1$), we obtain the usual additive entropy.
 
Let us now apply the  gravity-thermodynamics conjecture. 
This formulation states that the first law of thermodynamics can be applied to 
the cosmological horizon 
\cite{Jacobson:1995ab,Padmanabhan:2003gd,Padmanabhan:2009vy}.
The most commonly considered horizon is the apparent horizon 
\cite{Frolov:2002va,Cai:2005ra,Cai:2008gw}, which is defined as:
 \begin{equation}
\label{FRWapphor}
 {r_{A}}=\frac{1}{\sqrt{H^2+\frac{k}{a^2}}}.
\end{equation}
 To describe the thermodynamic properties of the horizon, one uses the 
temperature formula analogous to that of black holes, replacing the black hole 
event horizon with the apparent horizon \cite{Gibbons:1977mu}:
\begin{equation}
\label{Th}
T_h = \frac{1}{2\pi r_A}.
\end{equation}
Similarly, the entropy associated with the apparent horizon will be the 
standard Bekenstein-Hawking one, namely
\begin{equation}
\label{FRWHorentropy}
S_h=\frac{1}{4G} A.
\end{equation}
 By considering the energy flux across the apparent horizon, given by 
$\delta 
Q=-dE=A(\rho_m+p_m)H {r_A}dt$,
 and employing the first law of 
thermodynamics  $-dE=TdS$, alongside (\ref{Th}) and 
(\ref{FRWHorentropy}), we   derive  
\cite{Cai:2005ra}:
\begin{equation}
\label{FRWcFE1}
-4\pi G (\rho_m +p_m )= \dot{H} - \frac{k}{a^2}.
\end{equation}
\begin{equation}
\label{FRWcFE2}
\frac{8\pi G}{3}(\rho_m) =H^2+\frac{k}{a^2}-\frac{\Lambda}{3},
\end{equation}
with the integration constant giving rise to the cosmological constant.

 The gravity-thermodynamics conjecture has been applied   in the case of  
various extended entropies  
\cite{Lymperis:2018iuz,Arias:2019zug,Nojiri:2019skr,Geng:2019shx,
Saridakis:2020lrg, Lymperis:2021qty, 
  Sheykhi:2022gzb,DiGennaro:2022ykp,Luciano:2022ely, 
 Coker:2023yxr,Basilakos:2023kvk, 
 Salehi:2023zqg,Jizba:2023fkp,Jizba:2024klq,Ebrahimi:2024zrk,
Tyagi:2024cqp,Okcu:2024llu,Petronikolou:2024zcj,Mendoza-Martinez:2024gbx,
Yarahmadi:2024lzd,Tsilioukas:2025dmy}.
If one uses Tsallis entropy (\ref{Tsallisentropy}) instead of the standard one, 
then he finds \cite{Lymperis:2018iuz}
 \begin{eqnarray}
\label{FR1t}
&&H^2=\frac{8\pi G}{3}\left(\rho_m+\rho_{DE}\right)\\
&&\dot{H}=-4\pi G \left(\rho_m+p_m+\rho_{DE}+p_{DE}\right),
\label{FR2t}
\end{eqnarray}
where we have defined the effective dark energy density and pressure as 
 \begin{eqnarray}
&&
\!\!\!\!\!\!\!\!\!\!\!\!\!\!\!\!\!\!
\rho_{DE}=\frac{1}{8\pi G} 
\left\{  \Lambda +3H^2\left[1-\alpha \frac{ \delta}{ 2-\delta} 
H^{2(1-\delta) }
\right]
\right\},
\label{rhoDE1t}
\end{eqnarray}
\begin{eqnarray}
&& \!\!\!\!\!\!\!\!\!\!\!\!\!\!\!\!\!\!\!\!
p_{DE}=  \frac{1}{8\pi G}\left\{
\Lambda
+2\dot{H}\left[1-\alpha\delta H^{2(1-\delta)}
\right] 
+3H^2\left[1-\alpha\frac{\delta}{2-\delta}H^{
2(1-\delta)}
\right]
\right\},
\label{pDE1t}
\end{eqnarray}
where   
$\alpha\equiv (4\pi)^{\delta-1}\tilde{\alpha}$.
Lastly, the dark-energy equation-of-state   parameter  reads as
\begin{eqnarray}
w_{DE}\equiv\frac{p_{DE}}{\rho_{DE}}=-1-
\frac{     
  2\dot{H}\left[1-\alpha\delta H^{2(1-\delta)}
\right]
 }{\Lambda+3H^2\left[1-\frac{\alpha\delta}{2-\delta}H^{2(1-\delta)}
\right]}
\label{wDEt}.
\end{eqnarray}
As expected, for $\delta=1$ all above equations reduce to the ones of 
$\Lambda$CDM scenario.

\section{Correspondence between Myrzakulov $F(R,Q)$ gravity and Tsallis 
cosmology}
\label{Correspondence}

In this section, we investigate the conditions under which the Myrzakulov $ 
F(R,Q) $ gravity framework can be mapped to Tsallis cosmology.
We first study the background behavior, and then we proceed to the examination 
of perturbations.

\subsection{Background evolution}

To make the two theories coincide  at the background level, we 
appropriately choose the functions $ u(a, \dot{a}, \ddot{a}) $ and 
$ w(a, \dot{a}) $ in such a way that the modified Friedmann equations in both 
scenarios become identical. Observing the forms (\ref{rhoDEa1}),(\ref{pDEa1}) 
as well as (\ref{rhoDE1t}),(\ref{pDE1t}) we impose the ansatz 
\begin{equation}
    u(a, \dot{a}, \ddot{a}) =   \epsilon \left(\frac{\dot{a}}{a}\right)^\zeta+ 
\eta \left(\frac{\dot{a}}{a}\right)^\theta ,
\end{equation}
\begin{equation}
    w(a, \dot{a}) = \xi \left(\frac{\dot{a}}{a}\right) .
\end{equation}
Hence, we can see that  we obtain a 
correspondence between the two theories if we make the identifications  
$\lambda=3$, $\tilde{\Lambda} =2\Lambda$, 
$\eta=-12$, $\theta=2$, $\xi=10$, $\zeta=4-2\delta$ and $\epsilon=\frac{6 
\alpha \delta}{(2-\delta)(3-2\delta)}$.
 This correspondence may establish a   connection between metric-affine 
modifications of gravity and non-additive entropy formalisms in 
cosmology. 
 
Let us examine the cosmological evolution in more detail.   As the independent 
variable we  use the redshift $z$, defined  as  $ 
 1+z=1/a$ (we set the present scale factor to 1).
 Moreover, we  introduce the matter and dark energy density 
parameters 
  as
 \begin{eqnarray} \label{omatter}
&&\Omega_m=\frac{8\pi G}{3H^2} \rho_m\\
&& \label{ode}
\Omega_{DE}=\frac{8\pi G}{3H^2} \rho_{DE}.
 \end{eqnarray}  
Without loss of generality we will focus on dust matter, namely we take 
$p_m=0$. In this case the conservation equation  
gives   
 $\rho_{m} = \frac{\rho_{m0}}{a^3}$, with $\rho_{m0}$ the current value of the 
matter energy (from now on the subscript ``0" marks the present value of a 
quantity). Thus, for dust matter we have   
$\Omega_m=\Omega_
{m0} H_{0}
^2/(a^3 H^2)$. 
Moreover, we set the value of $\tilde{\Lambda}$ (and thus of $\Lambda$ too) in 
order 
for the Friedmann equation (\ref{FR1a}) applied at present time to give 
  $\Omega_{m0}\approx0.31$ in agreement with observations 
\cite{Planck:2018vyg}. Additionally, without loss of generality we set 
$\alpha=1$, namely to
its standard value.
 
\begin{figure}[!h]
\centering
\includegraphics[width=6.9cm]{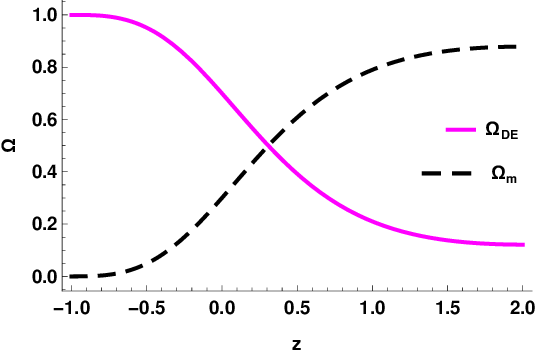} \\                                     
\includegraphics[width=6.9cm]{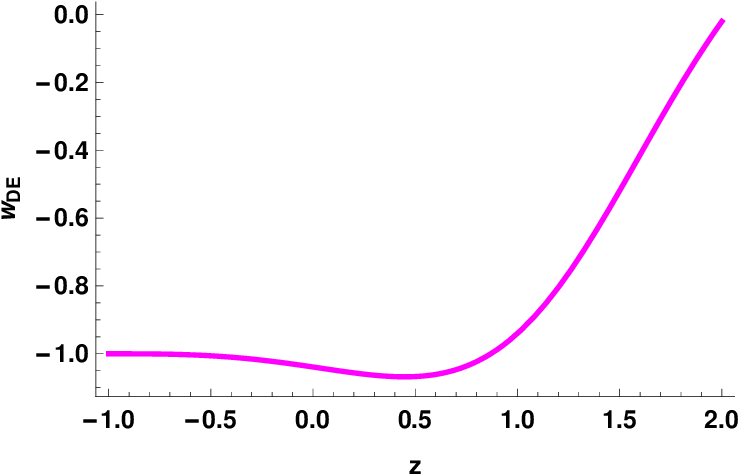} 
\caption{\it{Upper graph: The dark-energy density parameter $\Omega_{DE}$ 
(black-solid) and   the matter density parameter $\Omega_{m}$ 
(red-dashed), as a function of the redshift $z$,  for Tsallis cosmology with  
$\delta=1.1$ and     $\alpha=1$  in units where $H_0=1$,  and for Myrzakulov  
 gravity under the identifications mention in the text and for 
$\epsilon=\frac{6 
\alpha \delta}{(2-\delta)(3-2\delta)}=9.2$, 
in 
units where $H_0=1$. Lower graph: The  corresponding 
dark-energy 
equation-of-state parameter
$w_{DE}$.   In 
both graphs we have imposed    $\Omega_{m}(z=0)=\Omega_{m0}\approx0.31$. 
}}
\label{figback}
\end{figure}

In the upper panel of Fig.  \ref{figback} we present the evolution of the 
density parameters  $\Omega_{DE}(z)$ and $\Omega_{m}(z) 
= 1-\Omega_{DE}(z)$, for Myrzakulov cosmology and Tsallis cosmology, under the 
aforementioned identifications. In the lower panel we show  the   dark-energy 
equation-of-state parameter. Note that   we   extend  the evolution into the   
future too. As expected, there is no distinction between the two scenarios, 
since they were reconstructed to coincide at the background level.
As we can see, we obtain the sequence of matter and dark energy epochs, while  
in the future the universe results to the de-Sitter state. Concerning  the   
dark-energy 
equation-of-state parameter we can see that it may enter in the phantom regime, 
experiencing the phantom-divide crossing.

\begin{figure}[!h]
\centering
\includegraphics[width=12.cm]{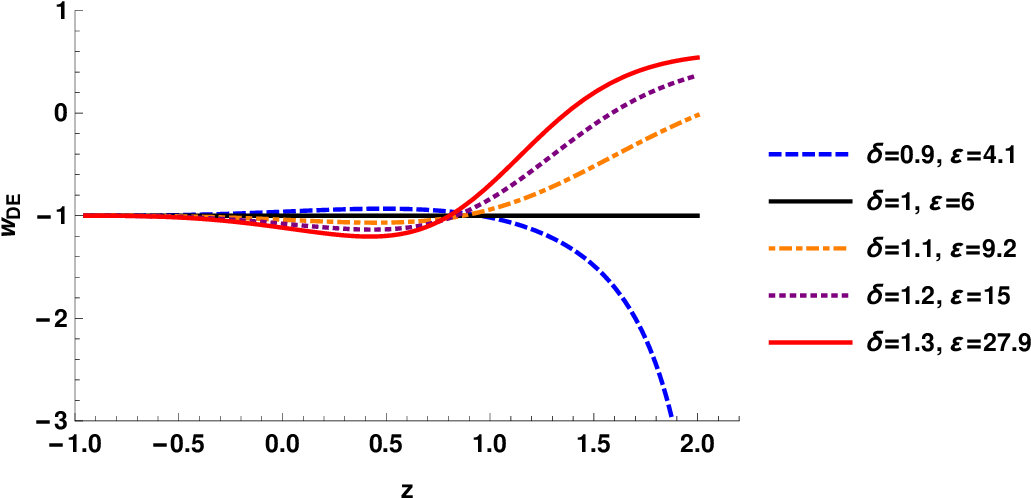}
\caption{\it{The   dark-energy 
equation-of-state parameter
$w_{DE}$  as a function of the redshift $z$,  for Tsallis cosmology with     
$\alpha=1$ in units where 
$H_0=1$, and various values 
of the parameter $\delta$, and for Myrzakulov gravity for
various parameters of $\epsilon=\frac{6 
\alpha \delta}{(2-\delta)(3-2\delta)}$.  For each curve we  set 
$\Omega_{m}(z=0)=\Omega_{m0}\approx0.31$ at present.}}
\label{figwde}
\end{figure}

Additionally, in Fig. \ref{figwde} we present the evolution of the dark-energy 
equation-of-state parameter for various values of the model parameters. 
Similarly to the previous figure, since $w_{DE}$ is a background-related 
quantity the two models coincide.
  
\subsection{Perturbation evolution}

Let us first review briefly the   behavior of Tsallis cosmology at the 
perturbative level. We introduce   the matter overdensity 
$\delta_m=\delta\rho_m/\rho_m$,   we focus  on dust matter 
and we set   $\alpha =1$ for convenience. 
Additionally, we assume that the 
dark-energy sector does not cluster. In this case we have  
\cite{Sheykhi:2022gzb}
 \begin{equation}
 \label{deltazt}
\delta^{\prime\prime}_{m}
+\frac{2\left (4\!-\!2\delta \right)-
\left(9\!-\!6\delta\! +\!8\pi G\Lambda H^{2\delta\! -\!4} \right )}{\left 
(4\!-\!2\delta \right)(1+z)}\delta^{\prime}_{m}
+
\frac{3^{\frac{1}{\delta -2}}\!
 \left [\left (1\!-\!2\delta 
\right)\rho^{\frac{1}{2-\delta}}_{m}-9\left(1\!-\!\delta \right)\Lambda  
\rho^{\frac{\delta -1}{2\!-\!\delta}}_{m}\right ]8\pi G 
}{2\left(2-\delta \right)^{2}H^{2}(1+z)^{2}}\delta_{m}=0,
\end{equation}
where  primes denote derivatives with respect to the redshift. As one can see, 
we obtain   a different friction term  as well as an effective  Newton  
constant  comparing to standard cosmology. 
In the case $\delta=1$  one 
obtains the standard result, which in the case of matter domination 
($\Omega_m\approx1$) yields
\begin{equation}
 \delta^{\prime\prime}_{m}+\frac{1}{2(1+z)}\delta^{\prime}_{m}
-\frac{3}{2(1+z)^2 } \delta_{m}=0,
\end{equation}
as expected.  

Let us now study the   perturbations in  Myrzakulov $ F(R,Q) $ gravity. 
Although the full analysis lies beyond the scope of the present work, we can 
obtain approximate expressions working within the mini-superspace framework. We 
will focus on the model where we have obtained a correspondence at the 
background level, and we assume non-clustering dark energy. We work   in the 
Newtonian gauge and we write the scalar perturbations of the metric as
\begin{equation}
ds^2 = -(1+2\Phi) dt^2 + a^2(1-2\Psi) \delta_{ij} dx^i dx^j.
\end{equation}
Inserting this into (\ref{miniLagr}), and imposing the aforementioned 
identifications that make the background behavior of the two theories to 
coincide, we find 
 \begin{equation}
 \label{deltaz}
\delta^{\prime\prime}_{m}
+\frac{2\left (4\!-\!2\delta \right)-
\left(9\!-\!6\delta\! +\!8\pi G\Lambda H^{2\delta\! -\!4} \right )}{\left 
(4\!-\!2\delta \right)(1+z)}\delta^{\prime}_{m}
+ \frac{ G\left(1 + \lambda \frac{w_{a} + w_{\dot{a}}}{2} + \frac{u_{a} 
+ u_{\dot{a}}}{2}\right)}{(1 + \lambda)(1+z)^2  }\delta_{m}=0.
\end{equation}

As we observe, although the friction term between the two theories is the same, 
the last term, related to the effective Newton constant is different. This 
is an important result since it implies that the growth of density 
perturbations and the matter power spectrum will be different in the two 
theories, and therefore confronting them with perturbation-related datasets, 
such as $\sigma_8$ and weak-lensing ones, will lead to different results. 
Hence, such an analysis will be useful in order to distinguish the two theories.

Let us proceed to a numerical example. We solve  (\ref{deltazt}) for Tsallis 
cosmology and  (\ref{deltaz}) for Myrzakulov gravity, under the aforementioned 
identifications in which the two models coincide. Then,   can calculate the 
  quantity  $
f\sigma_8(z)=f(z) \,\sigma(z)$,
 with  $f(z):=-\frac{d\ln\delta_m(z)}{d\ln z}$ and 
$\sigma(z):=\sigma_8\frac{\delta_m(z)}{\delta_m(0)}$ \cite{Abdalla:2022yfr}.
 
\begin{figure}[!h]
\centering
\includegraphics[width=7cm,height=10cm,angle=-90]{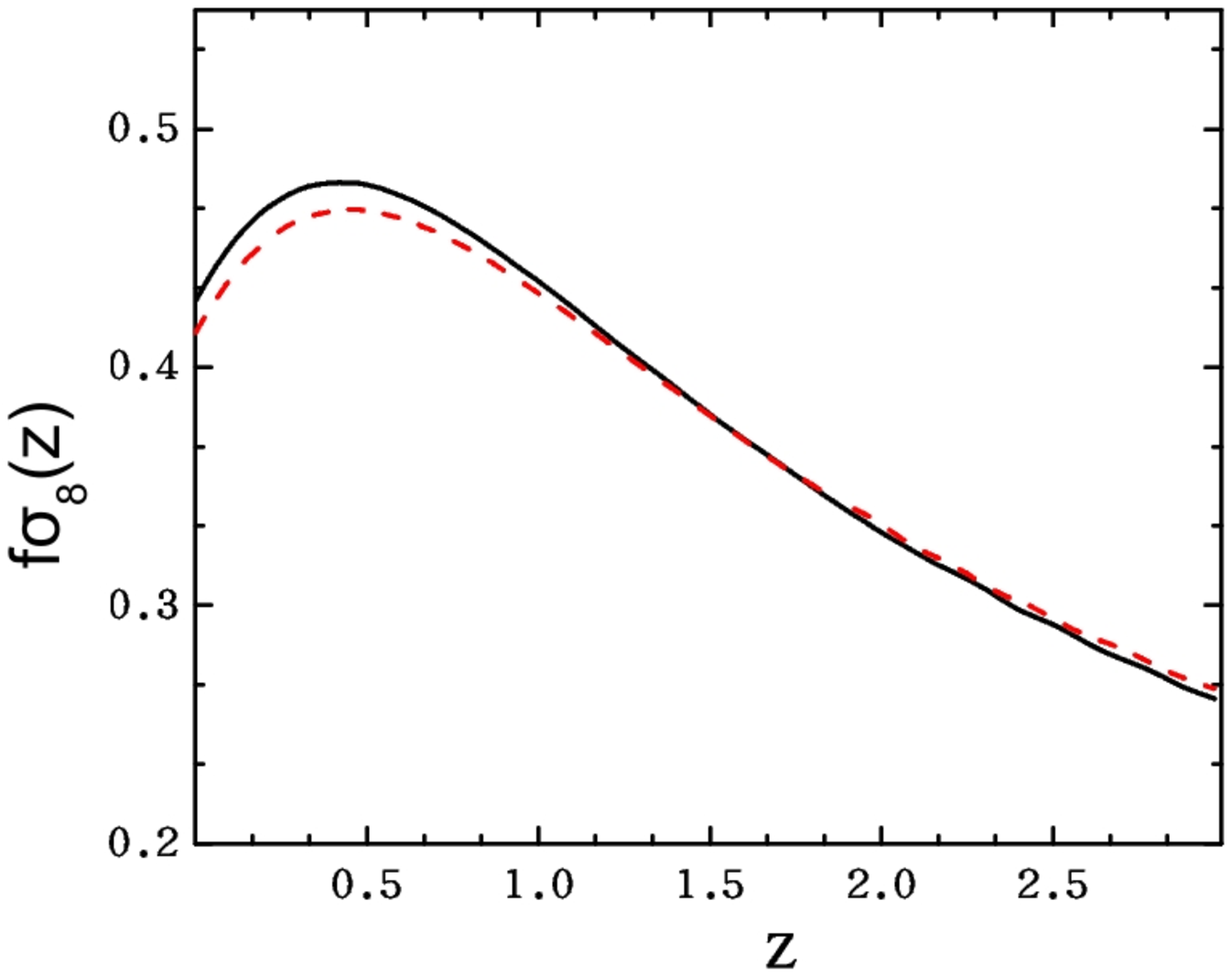}
\caption{{\it The behavior of of f$\sigma_8$  for Myrzakulov gravity (black 
solid) and  
Tsallis cosmology (red- dashed), under the aforementioned identifications and  
for $\delta=1.1$ and     $\alpha=1$, and   $\epsilon=\frac{6 
\alpha \delta}{(2-\delta)(3-2\delta)} =9.2$, in units where $H_0=1$, in 
which the two theories coincide at the background level. The curves 
do not coincide  and thus the behavior of perturbations can be used to 
distinguish the two theories.}}
\label{figpert}
\end{figure}

In Fig.   \ref{figpert}  we present the
evolution of $f\sigma_{8}$ for $\Lambda$CDM scenario, as well as for   Tsallis 
cosmology and Myrzakulov gravity, for the aforementioned identifications in 
which the two theories coincide at the background level. In this case the 
curves do not coincide anymore, and thus the behavior of perturbations can be 
used to distinguish the two theories.
 
\section{Conclusions}
\label{Conclusions}

Myrzakulov $F(R,Q)$ gravity is a modified gravitational framework that extends 
general relativity by incorporating both curvature and nonmetricity. By 
introducing an additional function of the Ricci scalar $R$ and the nonmetricity 
scalar $Q$, this theory provides a richer geometrical structure that allows for 
deviations from Einstein’s theory, and in particular in the context of 
cosmology. The inclusion of a non-special connection modifies the gravitational 
dynamics and affects the evolution of the universe, leading to novel 
phenomenological implications. Myrzakulov gravity has been   studied 
in various contexts, including inflation, late-time acceleration, and structure 
formation, demonstrating its potential as an alternative to the standard 
cosmological model.

On the other hand, Tsallis cosmology emerges from the gravity-thermodynamics 
conjecture, where the modified Friedmann equations arise from the first law of 
thermodynamics when employing Tsallis entropy instead of the usual 
Bekenstein-Hawking entropy. The Tsallis entropy formalism introduces a 
non-additive generalization of standard thermodynamic entropy, characterized by 
a parameter $\delta$ that encapsulates deviations from extensive statistical 
mechanics. Applying this approach to the cosmological setting results in 
modified equations that can describe an evolving dark energy sector. Hence, it 
  provides a thermodynamic interpretation of cosmic acceleration and 
can lead to interesting cosmological phenomenology.

In this work, we examined the correspondence between Myrzakulov $F(R,Q)$ 
gravity 
and Tsallis cosmology, focusing on their equivalence at the background level. 
By 
appropriately identifying the functional dependencies of the connection-related 
functions in Myrzakulov gravity, we reconstructed the effective dark energy 
sector of Tsallis cosmology. Our analysis 
demonstrated that both frameworks can give identical background evolution for 
specific function and parameter choices, reproducing the standard cosmological 
sequence of matter and dark energy domination. However, despite the fact that 
the two theories coincide   at the background level, their 
perturbation behavior exhibits  differences. In particular, the growth 
of density fluctuations and the effective Newton constant  deviate
between the two scenarios, indicating that perturbative observables, such as 
structure formation and weak-lensing ones, could serve as distinguishing 
factors between them.

Future research directions could study several extensions     of 
this correspondence. A more detailed investigation into the stability of 
perturbations in Myrzakulov gravity, particularly within the full metric-affine 
formalism, could offer more information   into the viability of the theory. 
Additionally, observational constraints from large-scale structure data, such 
as 
galaxy clustering and weak-lensing surveys, could be employed to quantify 
deviations from standard predictions. The extension of this correspondence to 
include other entropy-based modifications, such as Barrow and Kaniadakis 
entropies, may also provide a broader understanding of the connection between 
gravity and thermodynamics. These projects are left for future investigation.

\end{document}